\documentclass[preprint,preprintnumbers,amsmath,amssymb,nofootinbib]{revtex4}
\usepackage{bbm}
\usepackage{amsfonts}
\usepackage{booktabs}
\usepackage{mathrsfs}
\usepackage{epsfig}
\usepackage{graphicx}
\usepackage{dcolumn}
\usepackage{bm}
\usepackage{amsmath}
\usepackage{slashed}
\usepackage{subfigure}
\usepackage{multirow}
\usepackage{color}

\let\jnfont=\rm
\def\NPB#1,{{\jnfont Nucl.\ Phys.\ B }{\bf #1},}
\def\PLB#1,{{\jnfont Phys.\ Lett.\ B }{\bf #1},}
\def\EPJC#1,{{\jnfont Eur.\ Phys.\ Jour.\ C }{\bf #1},}
\def\PRD#1,{{\jnfont Phys.\ Rev.\ D }{\bf #1},}
\def\PRL#1,{{\jnfont Phys.\ Rev.\ Lett.\ }{\bf #1},}
\def\MPLA#1,{{\jnfont Mod.\ Phys.\ Lett.\ A }{\bf #1},}
\def\JPG#1,{{\jnfont J.\ Phys.\ G}{\bf #1},}
\def\CTP#1,{{\jnfont Commun.\ Theor.\ Phys.\ }{\bf #1},}
\def\ZPC#1,{{\jnfont Z.\ Phys.\ C }{\bf #1},}
\def\JHEP#1,{{\jnfont JHEP \ }{\bf #1},}
\def\Rv{\not{\hbox{\kern-1pt $R$}}}
\def\p{\not{\hbox{\kern-3pt $p$}}}

\begin{document}

\title{Anomalous Top-Higgs Couplings and Top Polarisation in Single Top and Higgs Associated Production at the LHC}

\author{Archil Kobakhidze\footnote{archilk@physics.usyd.edu.au}, Lei Wu\footnote{leiwu@physics.usyd.edu.au} and Jason Yue\footnote{j.yue@physics.usyd.edu.au}}
\affiliation{ ARC Centre of Excellence for Particle Physics at the Terascale, School of Physics, The University of Sydney, NSW 2006, Australia}%

\date{\today}

\begin{abstract}
In this paper, we put constraints on anomalous $\mathcal{CP}$-violating top-Higgs couplings using the currently available Higgs data and explore the prospect of measuring these couplings at 240 GeV TLEP. We find that the $\mathcal{CP}$-violating phase $\xi$ is currently limited in the range $|\xi|< 0.6\pi$ at 95\% C.L. and may be further constrained to $|\xi| <0.07\pi$ at TLEP. Under this consideration, we further investigate the observability of the scalar ($\xi =0$), pseudoscalar ($\xi =0.5\pi$) and mixed ($\xi =0.25\pi$)  top-Higgs interactions through the channel $pp \to t(\to \ell^+\nu_\ell b)h(\to b\overline{b})j$. We find that it is most promising to observe pure pseudoscalar interactions with $y_t=y_t^{SM}$, although this will be challenging due to a low signal to background ratio. We also find that the anomalous top-Higgs couplings can lead to sizeable differences in lepton forward-backward asymmetries and can be distinguished by measuring the lepton angular distributions from polarised top quarks at 14 TeV LHC.
\end{abstract}

\maketitle


\section{\label{sec:level1}Introduction}
The discovery of the Higgs boson at the Large Hadron Collider (LHC) \cite{higgs-atlas,higgs-cms} is a major step towards elucidating the electroweak symmetry breaking (EWSB) mechanism. To ultimately establish its nature, a precise measurement of the Higgs couplings to fermions and gauge bosons, and Higgs self-coupling is an important task of experiments at LHC and future colliders \cite{review,hf}. Although the observed Higgs signals are still plagued with large uncertainties, they have at least firmly established the key Higgs production and decay channels predicted by the Standard Model (SM). In turn, the Higgs main production channel $gg\rightarrow h$ indirectly indicates the existence of top-Higgs interactions.

In the SM, top quark is the heaviest fermion and hence has the strongest coupling to the Higgs boson. As such, it plays an important role in the EWSB and in various cosmological phenomena, such as electroweak phase transition and baryogenesis. The associated production of the top pair with Higgs boson have been widely investigated at the LHC \cite{tth_old,tth_new} as a direct probe of the top-Higgs Yukawa coupling. Based on an integrated luminosity of $L= 20.3$ fb$^{-1}$, the ATLAS collaboration has analysed $pp \to t\overline{t}h$, with $h\to b\overline{b}$, and set a $95\%$ C.L. limit \cite{atlas_tth} on the $t\overline{t}h$ cross section, $\sigma_{t\overline{t}h}<4.1\sigma^{SM}_{t\overline{t}h}$. The determination of the dominant background, $t\overline{t}\ +$ jets, is expected to be improved by the copious production of top quarks at the LHC \cite{ttj}. However, even if the $t\overline{t}h$ coupling can be measured with sufficient accuracy, information on the relative phase between the top-Higgs Yukawa coupling and gauge-Higgs coupling will still be lacking.

In this regard, the search for Higgs boson production in association with a single top have proposed in Refs.~\cite{thj_old}. Like single top productions, there are three different production modes characterised by the virtuality of the $W$ boson \cite{maltoni_thj1}. The $t$ channel process $pp \to thj$ with a space-like $W$ has the largest cross section amongst these production mode, reaching $\sim88.2$ fb (14 TeV) at NLO QCD in the SM \cite{maltoni_thj2}. The most important feature of $thj$ production is that the interference between the contributing processes with $ht\overline{t}$ and $hWW$ couplings allows direct examination both the modulus ($y_t$) and the $\mathcal{CP}$ violating phase ($\xi$) of the top-Higgs Yukawa coupling \cite{maltoni_thj1,maltoni_thj2,biswas_thj,ellis_thj,englert,lee_thj}. Such anomalous top-Higgs couplings may result from various new physics models \cite{eff,cpv_model1,cpv_model2,cpv_model3,cpv_model4,archil}. The CMS collaboration has very recently presented the result on $thj$ searches in the $h \to \gamma\gamma$ channel, and obtained a weak bound on the cross section of events with inverted top-Higgs coupling \cite{cms_thj}. An equally important feature is that the top quark produced via the weak  interaction in $thj$ is left-handed in the SM. It is therefore expected that non-standard top-Higgs couplings will affect the polarisation states of top quark and change the angular distributions of the top quark decay products \cite{ellis_thj}. Such a polarisation asymmetry has been widely used to probe the anomalous top quark interactions at the LHC  \cite{top_pol}. The precise measurement of $thj$ channel opens a new window to probe the top quark Yukawa couplings and new physics at the LHC.

In this work, we examine the current and future constraints on the $\mathcal{CP}$-violating $t\overline{t}h$ couplings based on present LHC data and expected 240GeV TLEP sensitivity respectively.  We investigate the observability of $pp \to thj$ with $h \to b\overline{b}$ for the scalar, pseudoscalar and mixed interactions of top-Higgs. The potential to discriminate such anomalous top-Higgs coupling is studied by performing reconstructed-level Monte Carlo simulations at 14 TeV HL-LHC. This paper is organised as follows: in section 2, we present the Higgs data constraints on the anomalous top quark Yukawa couplings; in section 3, we discuss their observability by analysing $pp \to thj$ production; conclusions are drawn in section 4.

\section{Higgs data Constraints}
In some new physics models, the top quark Yukawa coupling can be different from the SM prediction. The new physics effects on $t\overline{t}h$ coupling can be parameterised by a minimal set of the gauge invariant dimension-six operators \cite{eff}. The most general Lagrangian of the $t\overline{t}h$ interaction in the broken phase can be parameterised as follows:
\begin{equation}
{\cal L} \supset -\frac{y_t}{\sqrt{2}}\overline{t}(\cos\xi+i\gamma^{5}\sin\xi)th,   \qquad \qquad  \xi \in (-\pi,\pi]
\end{equation}
where $y_t$ takes the value $y^{SM}_t=\sqrt{2}m_t/v$ and $\xi =0$ in the SM, with $v\approx 246$ GeV being the vacuum expectation value of the Higgs field. It is useful to define the scalar and pseudoscalar components of the anomalous top-Higgs interaction normalised to the tree-level  SM coupling as  $C_S= y_t \cos\xi/y^{SM}_t$ and $C_P= y_t \sin\xi/y^{SM}_t$ respectively. It should be noted that such $\mathcal{CP}$-violating interactions contribute to the electric dipole moment (EDM). However, the bounds on the coupling $C_P$ depend on the assumption of Higgs couplings to other light fermions \cite{top_cp_review,zupan}. Since these couplings are practically unobservable at the LHC, we do not impose EDM constraints in this work. Other constraints from low-energy physics observables, such as $B_s-\overline{B}_s$ and $B \to X_s \gamma$, remain relatively weak \cite{zupan}.

\begin{figure}[h]
\centering
\includegraphics[width=.45\textwidth,clip=true,trim=7mm 8mm 10mm 16mm]{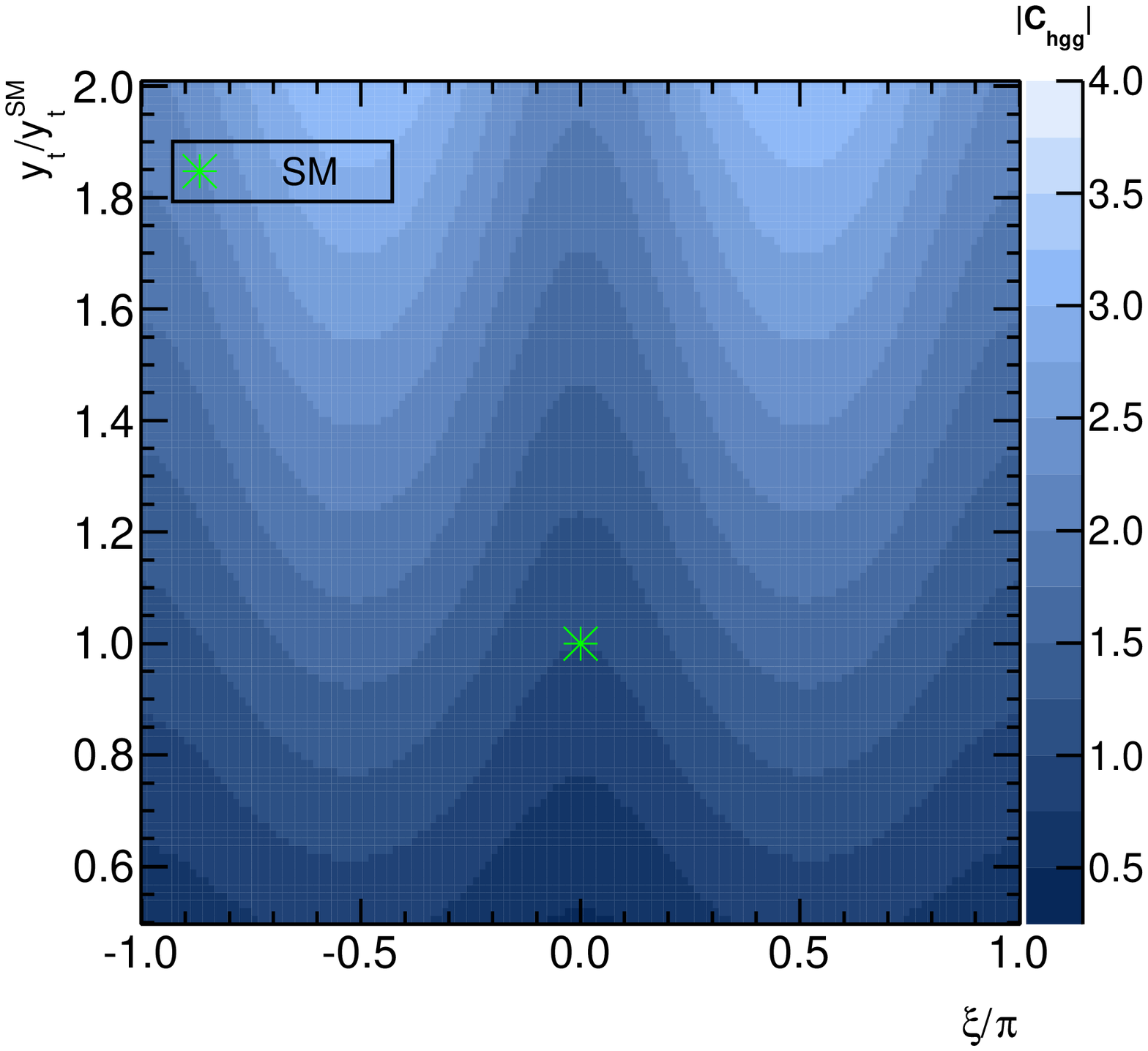}
\includegraphics[width=.45\textwidth,clip=true,trim=7mm 8mm 10mm 16mm]{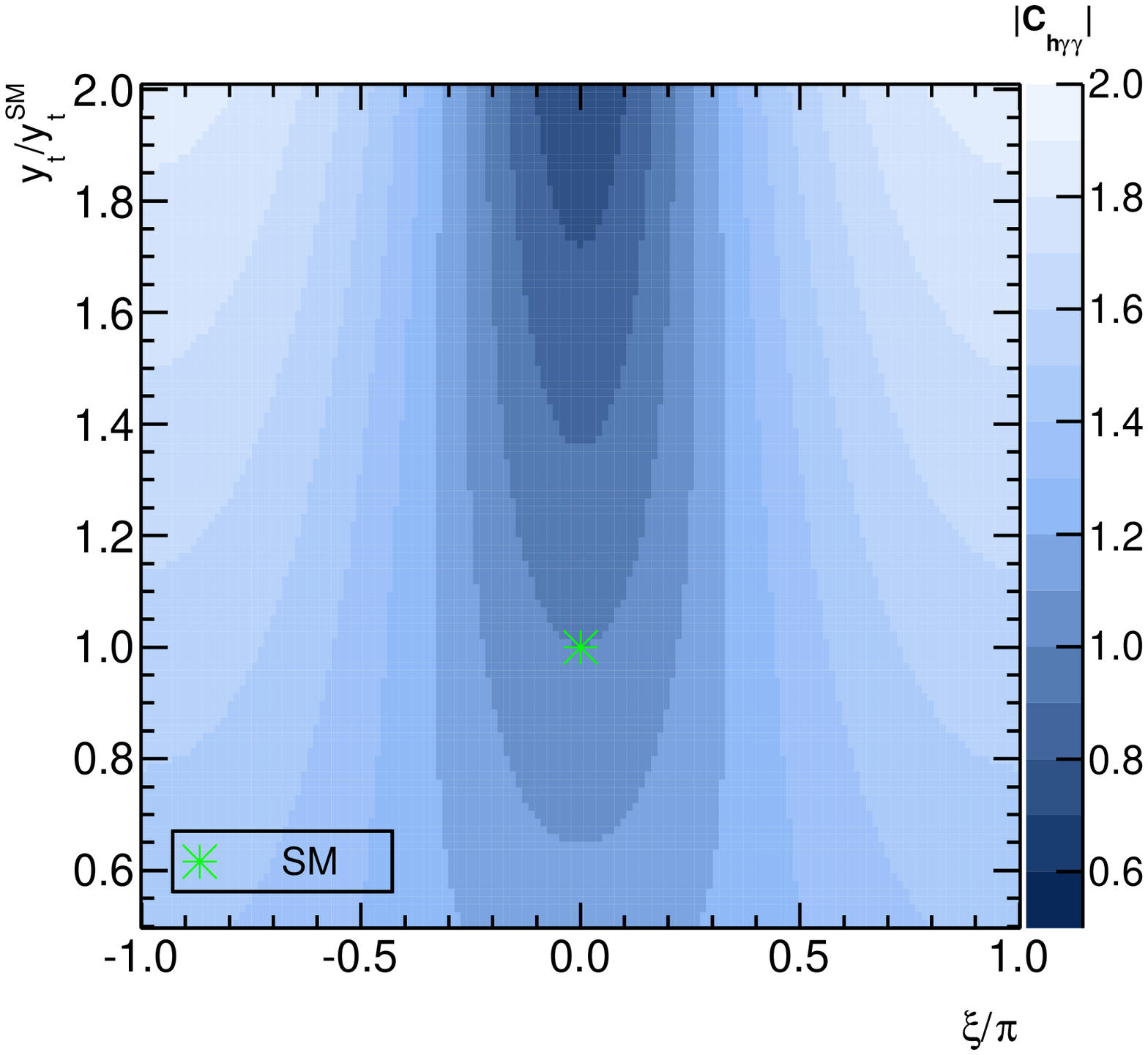}\vspace{-0.5cm}
\caption{Reduced couplings $C_{hgg}$ and $C_{h\gamma\gamma }$ as a function of $y_t$ and $\xi$.}
\label{red_coup}
\end{figure}

An obvious consequence of non-standard top-Higgs interaction is that the production rate of $gg \to h$ and decay width of $h \to \gamma\gamma$ will deviate from those in the SM. In Fig.~\ref{red_coup}, we show the dependence of the reduced couplings $C_{hgg}$ and $C_{h\gamma\gamma }$ in terms of $y_t$ and $\xi$. It could be seen that the coupling $C_{hgg}$ at $\xi=\pm \pi/2$ is larger than that at $\xi=0$. The reason is that the pseudoscalar interaction of $t\overline{t}h$ can lead to a larger form factor than the scalar interaction in the calculation of $gg \to h$. We can also notice that when $y_t$ increases, the reduced coupling $C_{hgg}$ becomes larger. However, this is different from the case in the coupling $C_{h\gamma\gamma}$, where the larger $y_t$ will induce a stronger cancellation between top quark loop and $W$ boson loop. 

\begin{figure}[h!]
\centering
\includegraphics[width=.45\textwidth,clip=true,trim=7mm 8mm 12mm 16mm]{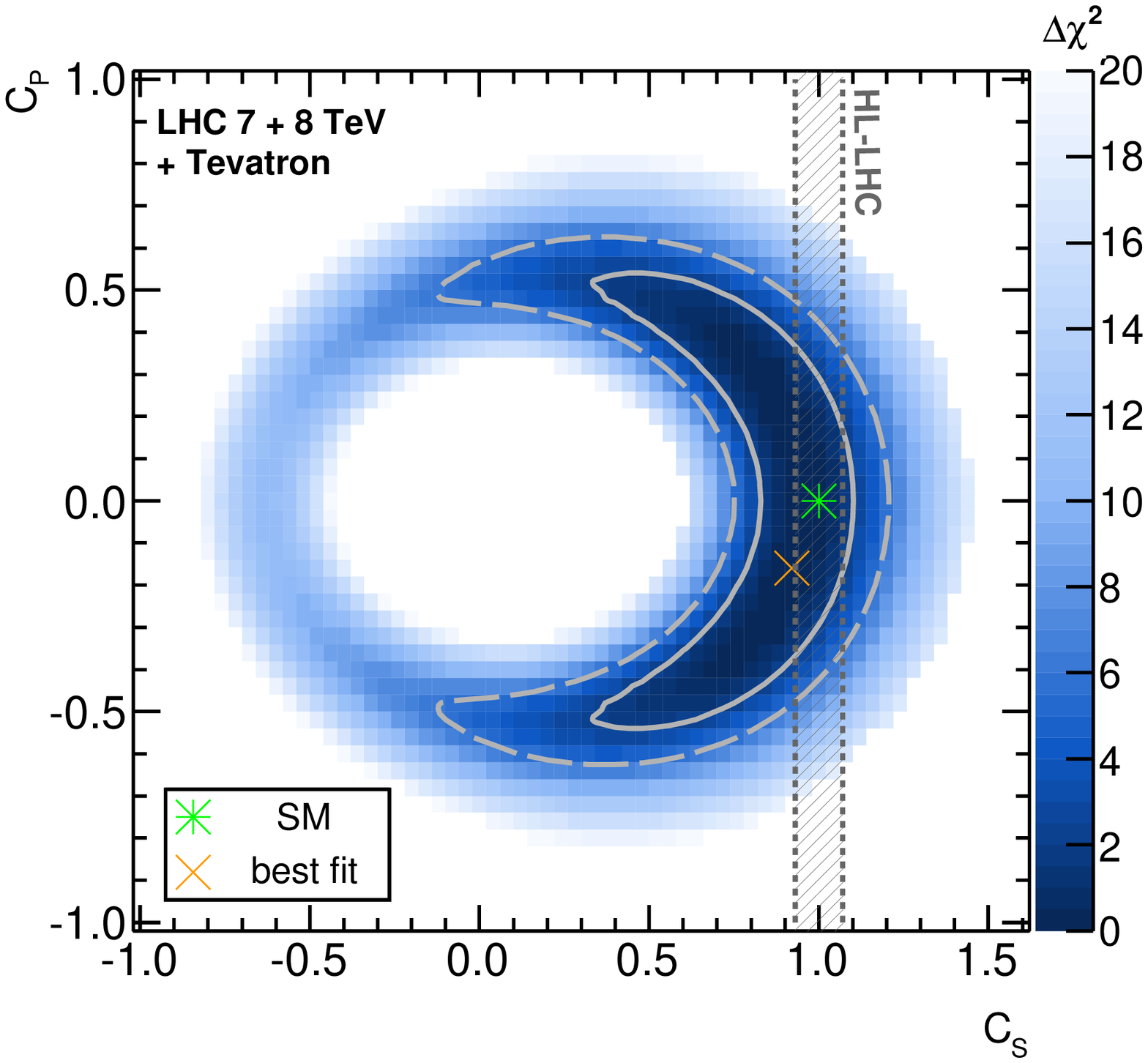}
\includegraphics[width=.45\textwidth,clip=true,trim=7mm 8mm 12mm 16mm]{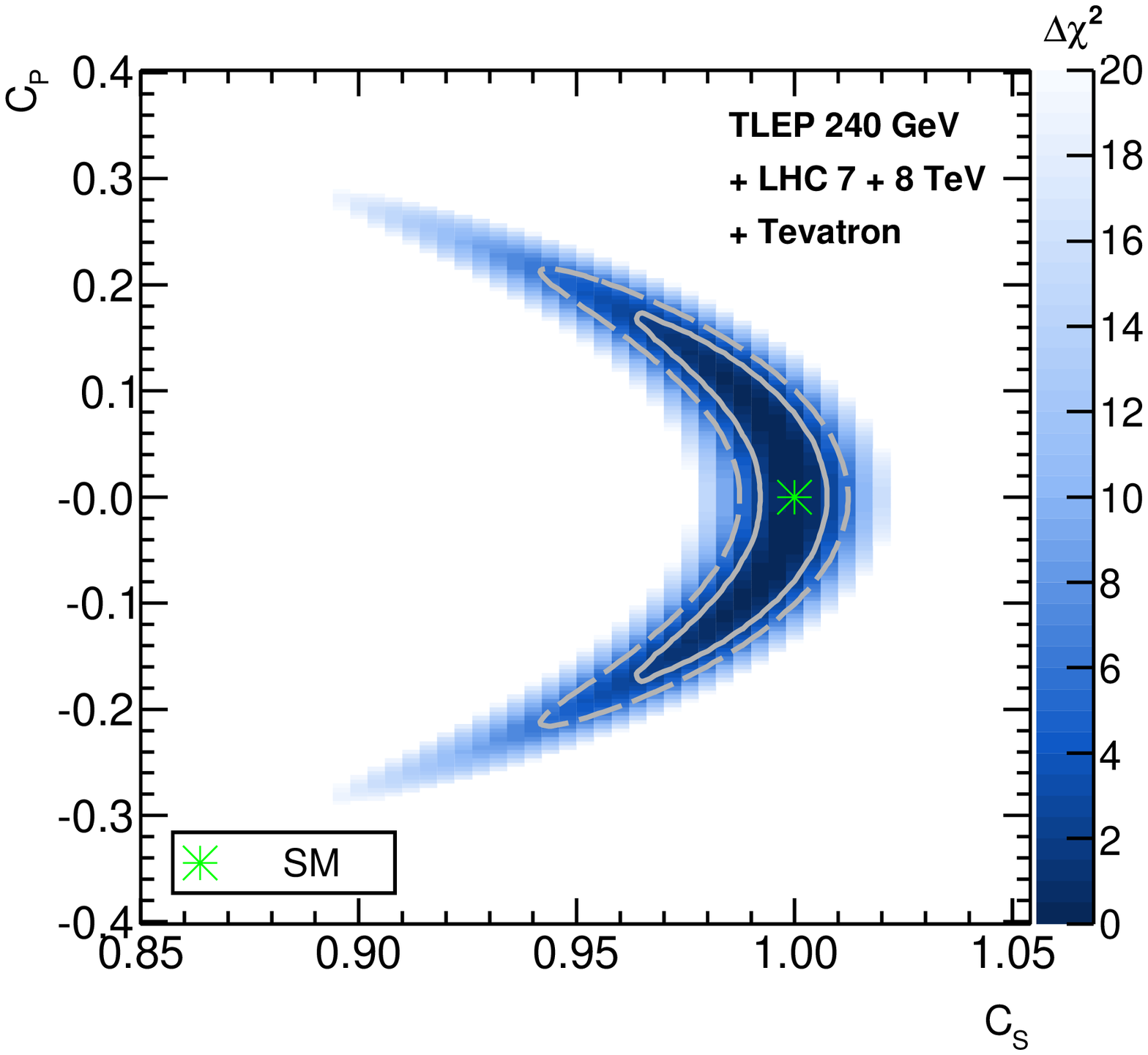}\vspace{-0.5cm}
\caption{The Higgs data constraints on the anomalous couplings $C_S$ and $C_P$ at the LHC and the expected sensitivity of these couplings at 240 GeV TLEP. The solid and dashed lines correspond to 68\% and 95\% C.L. respectively. The shadowed region represents the expected measurement uncertainty at HL-LHC.}
\label{cpv}
\end{figure}

We impose the latest Higgs data constraints on the anomalous couplings $C_S$ and $C_P$ by calculating $\chi^2$ with the public package \textsf{HiggsSignals-1.2.0} \cite{higgssignals}, which includes all the available data sets from the ATLAS, CMS, CDF and D{\O}  collaborations. In Fig.~\ref{cpv}, we present the Higgs data constraints on the anomalous couplings $C_S$ and $C_P$ using current Higgs data and the prospect of improving the bounds at TLEP with $\sqrt{s}=240$ GeV. We find that reduced pseudoscalar couplings in the range  $|C_P|>0.6$ have been excluded at 95\% C.L by the current Higgs data and that positive scalar couplings $C_S>0.5$ are strongly favoured, consistent with Ref.~\cite{ellis_thj,Maito_fit,Nishiwaki_fit}. As the $\mathcal{CP}$-phase $\xi$ increases from 0 to $\pi/2$, the 95\% C.L. allowed region for $y_t/y_t^{SM}$ reduces from $0.7$ -- $1.2$ to $0.4$ -- $0.6$. The expected measurement of $C_S$ at 14 TeV HL-LHC (3000 fb$^{-1}$) will further  constrain $C_P$ to the range $|C_P|<0.4$.
To estimate the bounds at TLEP, all measured Higgs couplings  are assumed to be the same as the SM predictions, and the expected measurement uncertainties are taken from Table 1-16 of Ref.~\cite{snowmass}. The main constraints come from the precise determinations of loop induced, reduced $C_{hgg}$ and $C_{h\gamma\gamma}$ couplings at TLEP. From Fig.~\ref{cpv}, we see that the allowed range of $C_P$ at 95\% C.L. shrinks to $|C_P|<0.2$, and $C_S$ is very close to one.

\begin{figure}[h!]
\centering
\includegraphics[width=.45\textwidth,clip=true,trim=4mm 8mm 12mm 16mm]{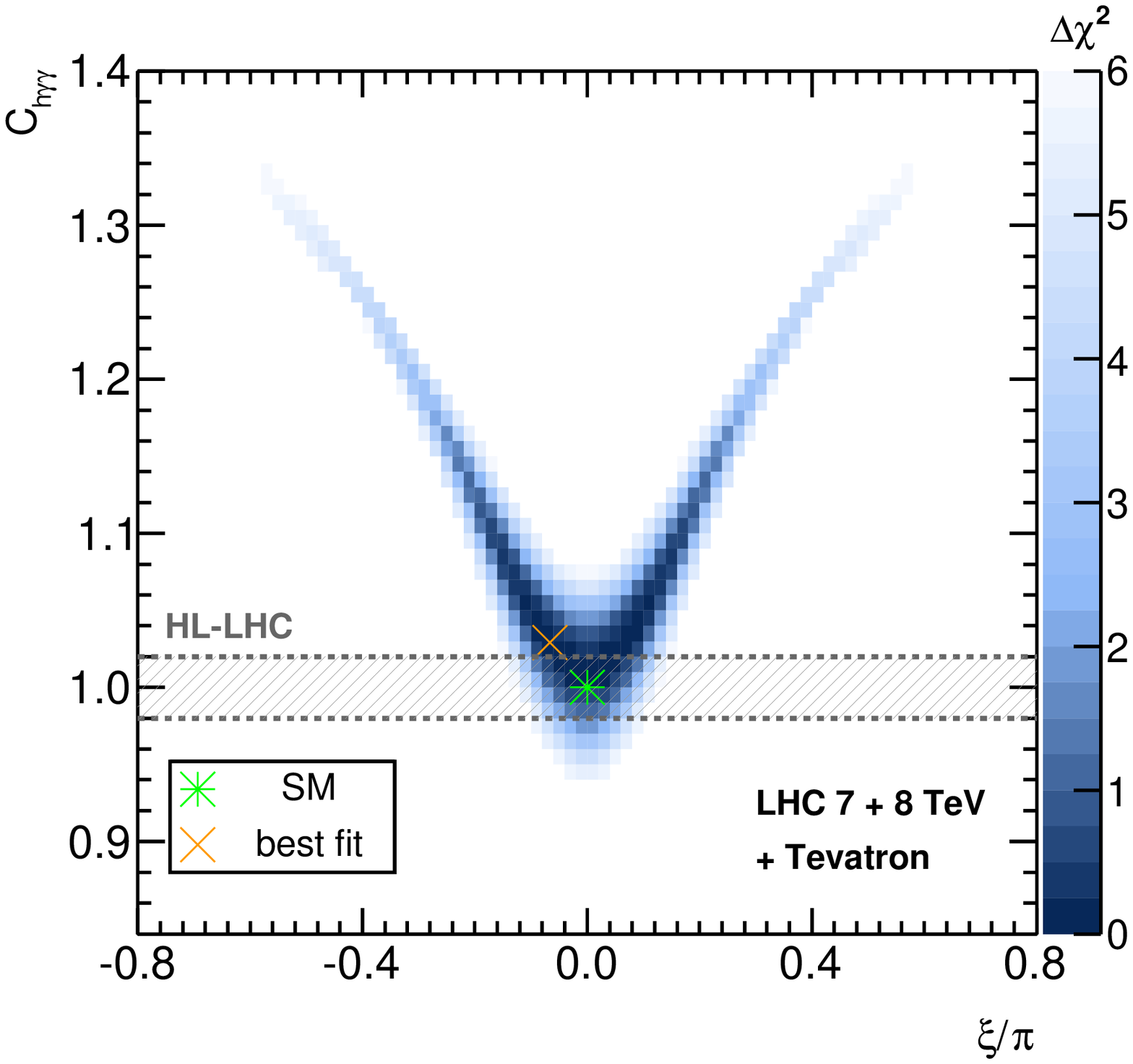}
\includegraphics[width=.45\textwidth,clip=true,trim=4mm 8mm 12mm 16mm]{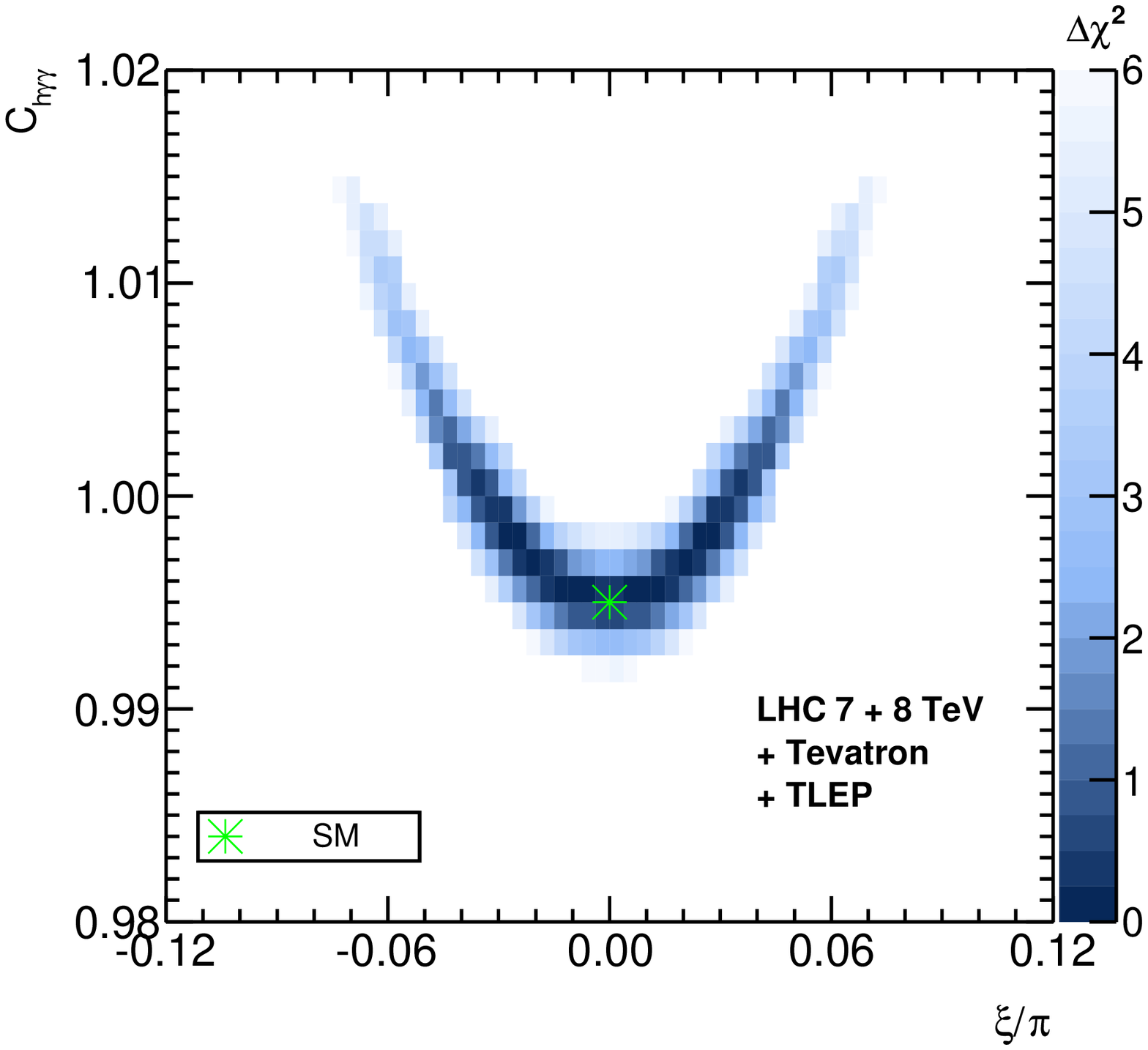}\vspace{-0.5cm}
\caption{Left: The Higgs diphoton reduced coupling $C_{h\gamma\gamma}$ versus $\mathcal{CP}$-violating phase $\xi$  in the 95\% C.L  allowed range in Fig.~\ref{cpv}. The shaded region is the expected measurement uncertainty of $C_{h\gamma\gamma}$ at HL-LHC; Right: The expected constraints on $C_{h\gamma\gamma}$ and $\xi$ at 240 GeV TLEP.}

\label{couplings}
\end{figure}

In Fig.~\ref{couplings}, we project the samples in the above 95\% C.L. range allowed by the current Higgs data on the plane of the Higgs-diphoton reduced coupling $C_{h\gamma\gamma}$ versus the $\mathcal{CP}$ phase $\xi$. From Fig.~\ref{couplings}, we see that the effective coupling $C_{h\gamma\gamma}$ can be sizeably enhanced by the constructive contribution of the top-quark loop with $\mathcal{CP}$-violating couplings, reaching a maximal value of $\sim$1.32. The bound on $\xi$ is expected to improve from $|\xi|<0.6\pi$ at the LHC to $|\xi|<0.07\pi$ at 240 GeV TLEP.

\section{anomalous $t\overline{t}h$ couplings and $thj$ production}
In the numerical calculations we take the SM input parameters as follows \cite{pdg}:
\begin{align}
m_t = 173.07{\rm ~GeV},\quad &m_{Z} =91.1876 {\rm ~GeV}, \quad \alpha(m_Z) = 1/127.9, \\ \nonumber
\sin^{2}\theta_W = 0.231,\quad &m_h =125 {\rm ~GeV}, \quad \alpha_{s}(m_Z)=0.1185.
\end{align}
By performing the Monte Carlo simulation, we investigate the observability of the anomalous top-Higgs couplings through the single top and Higgs associated production at the LHC
\begin{eqnarray}\label{process}
pp \to thj \to b \ell^{+} \nu b \overline{b} j,
\end{eqnarray}
where $j$ denotes the light jets and $\ell^{+} = e^{+},\mu^{+}$. Our signal is characterised by multi-jets (1 forward jet + 3 $b$-jets) + 1 lepton + missing energy (due to the neutrinos) in the final states. Although the $h \to b\overline{b}$ decay mode suffers a loss of efficiency in Higgs mass reconstruction, this shortcoming is mildly compensated by the large branching ratio of $h \to b\overline{b}$. Such a signature resembles the $t\overline{t}h$ topology analysed by ATLAS and CMS collaborations at the LHC Run-I, where at least 3 $b$-jets are required \cite{tth_bb}.

\begin{figure}[h]
\centering
\includegraphics[width=.5\textwidth,clip=true,trim=4mm 8mm 14mm 14mm]{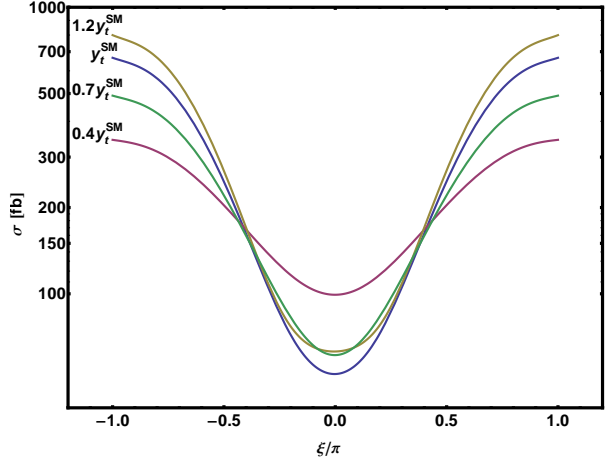}
\caption{$pp\rightarrow thj$ production cross section (fb) as a function of $\mathcal{CP}$ phase $\xi$ at 14 TeV LHC. The conjugated process $\overline{t}hj$ is not included here.}
\label{Xsection}
\end{figure}

We implement the $\mathcal{CP}$-violating interaction of $t\overline{t}h$ in (\ref{cpv}) by using the package \textsf{FeynRules} \cite{feynrules} and generate the parton-level signal and background events with \textsf{MadGraph5} \cite{mad5}. In Fig.~\ref{Xsection}, we show the resulting the $thj$ production cross section with no cuts on the final state kinematic distribution,  as a function of $\xi$ for a selection of $y_t$ values. The values $y_t=0.7y^{SM}_{t}$ and $y_t=1.2y^{SM}_{t}$ respectively correspond to the 95\% C.L. lower and upper bounds of $y_t$ for $\xi=0$, whilst $y_t=0.4y^{SM}_{t}$ is the lower bound for $\xi=\pi/2$. We can see that the maximal value of the cross section occurs at  $|\xi|=\pi$, due to the constructive interference of the contributions involving the $hWW$ and $ht\overline{t}$ couplings.

Parton showering and fast detector simulations are subsequently performed by \textsf{PYTHIA} \cite{pythia} and \textsf{Delphes} \cite{delphes}. Jets are clustered by using the anti-$k_t$ algorithm with a cone radius $\Delta R=0.7$ \cite{anti-kt}. We keep the default cuts setting when generating the parton level events and set both the renormalisation  ($\mu_R$) and factorisation ($\mu_F$) scale to the default event-by-event value and take CTEQ6L as the parton distribution function \cite{cteq}. We adopt the $b$-jet tagging efficiency ($\epsilon_b$) formula \cite{cms-b} that is a function of the transverse momentum and rapidity of the jets, with $\epsilon_b =0$ in the forward region ($|\eta|>2.5$). We also include a misidentification probability of  10\% and 1\% for $c$-jets and light jets respectively. The mis-tag of QCD jets is assumed to be the default value as in \textsf{Delphes}. The number of events generated for both  the signals and backgrounds  in our calculations is $1.2\times 10^6.$

The main SM backgrounds are: (B1) $pp \to t\overline{t}$, which can fake the signal when one light jet from the (anti-)top quark hadronic decay is misidentified as a $b$-jet. (B2) $pp \to tZ(\to b\overline{b})j$, which is an irreducible background but with a pair of $b$-jets coming from $Z$ boson; and (B3) the irreducible QCD process $pp \to tb\overline{b}j$. Since the last two backgrounds have been demonstrated to be small \cite{lee_thj,maltoni_thj2}, we will focus on background (B1). To include the QCD effects, we generate parton-level events of $t\bar{t}$ with up to two jets that are matched to the parton shower using the MLM-scheme \cite{Mangano:2006rw} with merging scale $x_q = 20$ GeV.
We impose the basic cuts on the final states as follows:
\begin{eqnarray}\label{basic}
  \Delta R_{ij} > 0.4\ ,& \quad  i,j &= b,j \ \text{or}\  \ell \\ \nonumber
  p_{T}^b > 25 \ \text{GeV},&\quad |\eta_b| &<2.5 \\ \nonumber
 p_{T}^\ell > 25 \ \text{GeV},& \quad  |\eta_\ell|&<2.5  \\ \nonumber
 p_{T}^j > 25 \ \text{GeV},& \quad  |\eta_j|&<4.7.
\end{eqnarray}

\begin{figure}[h]
\centering
\includegraphics[width=.45\textwidth,clip=true,trim=6mm 6mm 10mm 10mm]{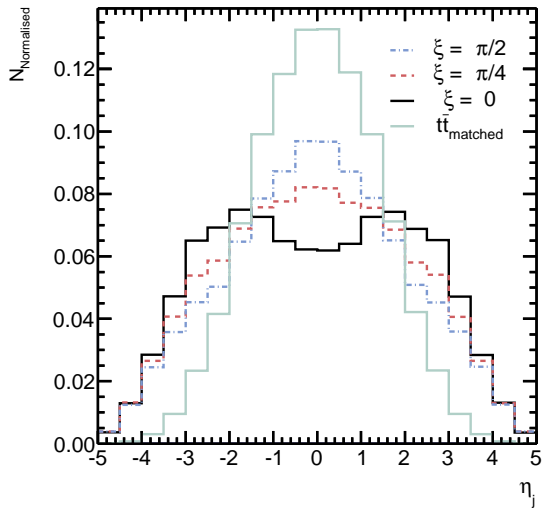}\vspace{-0.5cm}
\caption{The pseudorapidity distributions of the leading jet in the signals and backgrounds.}
\label{eta}
\end{figure}
In Fig.~\ref{eta}, we plot the pseudorapidity distributions of the leading jet in the signals and backgrounds. It can be observed that most $t\overline{t}$ events of have a leading jet in the central region, which  differs significantly from the signal, where a forward spectator jet  accompanies the top quark and Higgs boson. The pseudorapidity of the leading jet is therefore required to satisfy $2.5<|\eta_{j_1}|<4.7$ in order to reduce the $t\overline{t}$ background.

\begin{figure}[h!]
\centering
\includegraphics[width=.45\textwidth,clip=true,trim=6mm 2mm 10mm 10mm]{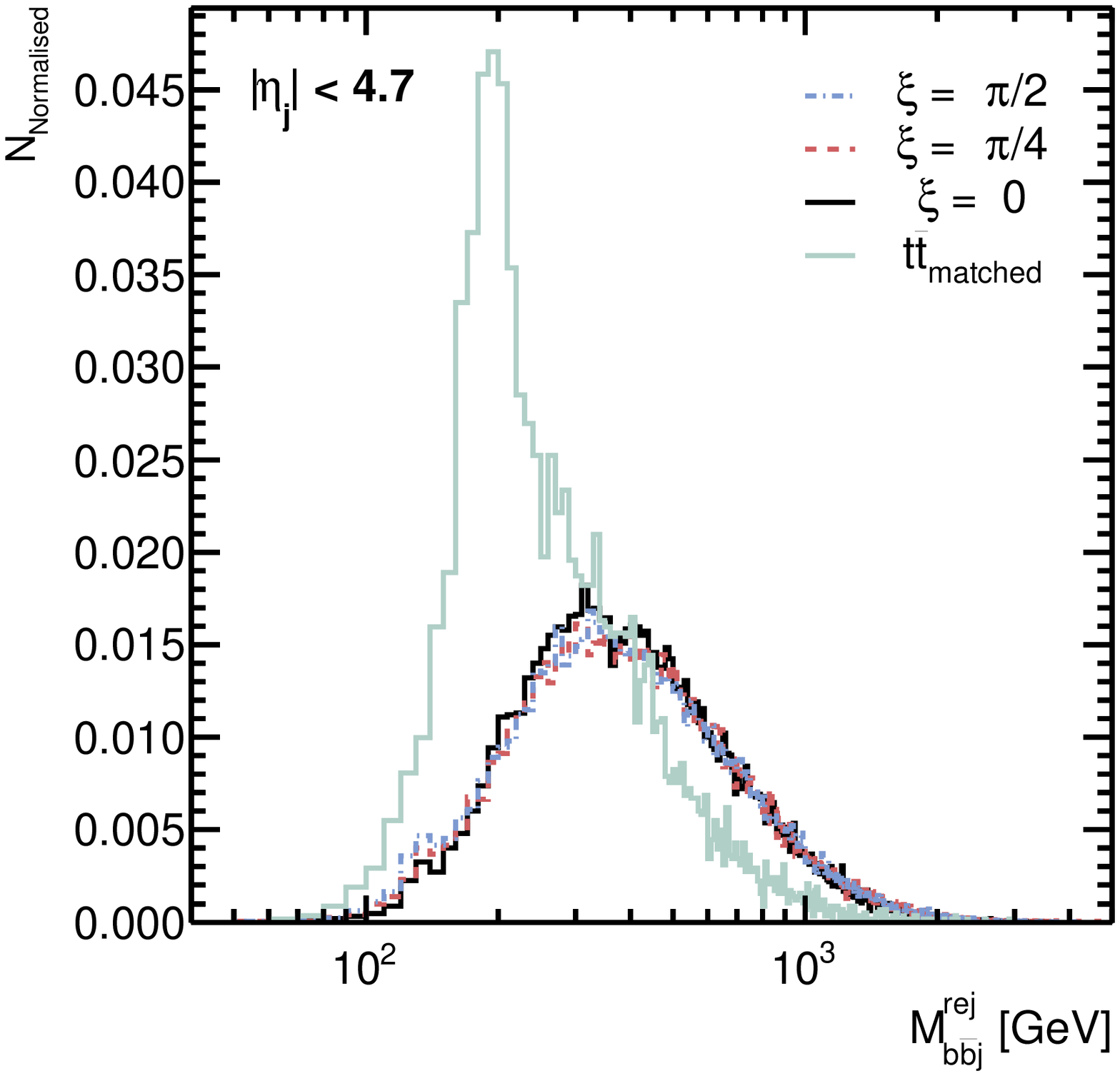}
\includegraphics[width=.45\textwidth,clip=true,trim=6mm 2mm 10mm 10mm]{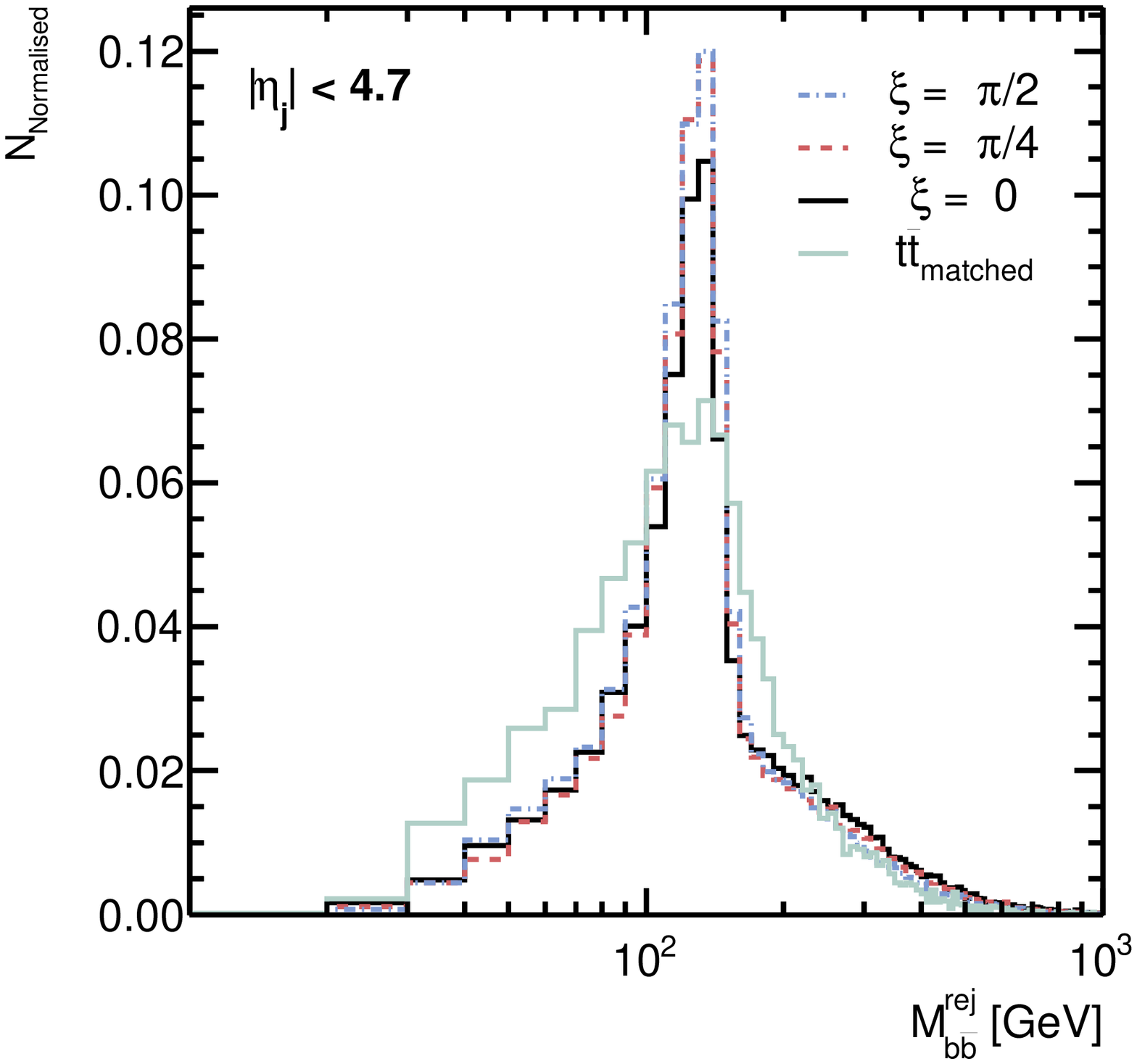}
\vspace{-0.5cm}
\caption{The normalised invariant mass distributions of $M^{rec}_{b\overline{b}}$ and $M^{rec}_{b\overline{b}j}$ with only basic cuts. Here the two $b$-jets are those rejected by the minimisation of $M_{b\ell}$.}
\label{before}
\end{figure}
Another cut which further suppresses the $t\overline{t}$ background is the invariant mass cut on the two $b$-jets from the Higgs boson decay. As proposed in Ref.~\cite{lee_thj}, the $b$-jet from the top quark decay can be tagged by selecting the minimal one among the invariant masses of each $b$-jet and the lepton, which should also satisfy $M^{min}_{b\ell}<200$ GeV. The remaining two $b$-jets are the considered as possible daughters of the Higgs boson for the signal. Given that the third $b$-jet in the $t\overline{t}$ background comes mainly from the misidentification of the jets in the top quark hadronic decay, the invariant mass $M^{rej}_{b\overline{b}j}$ and the two $b$-jets rejected by the minimisation of $M_{b\ell}$ should have a peak around $m_t$ for $t\overline{t}$, as is evident in Fig.~\ref{before}. We further find that the signal $M^{rej}_{b\overline{b}}$  invariant mass peaks are more narrow than those of the backgrounds but are still relatively broad around the Higgs mass, reducing the effectiveness of the cut $|M^{rej}_{b\overline{b}}-m_h|<15$ GeV in enhancing the observability of the signals.

\begin{figure}[h!]
\centering
\includegraphics[width=.45\textwidth,clip=true,trim=6mm 2mm 10mm 10mm]{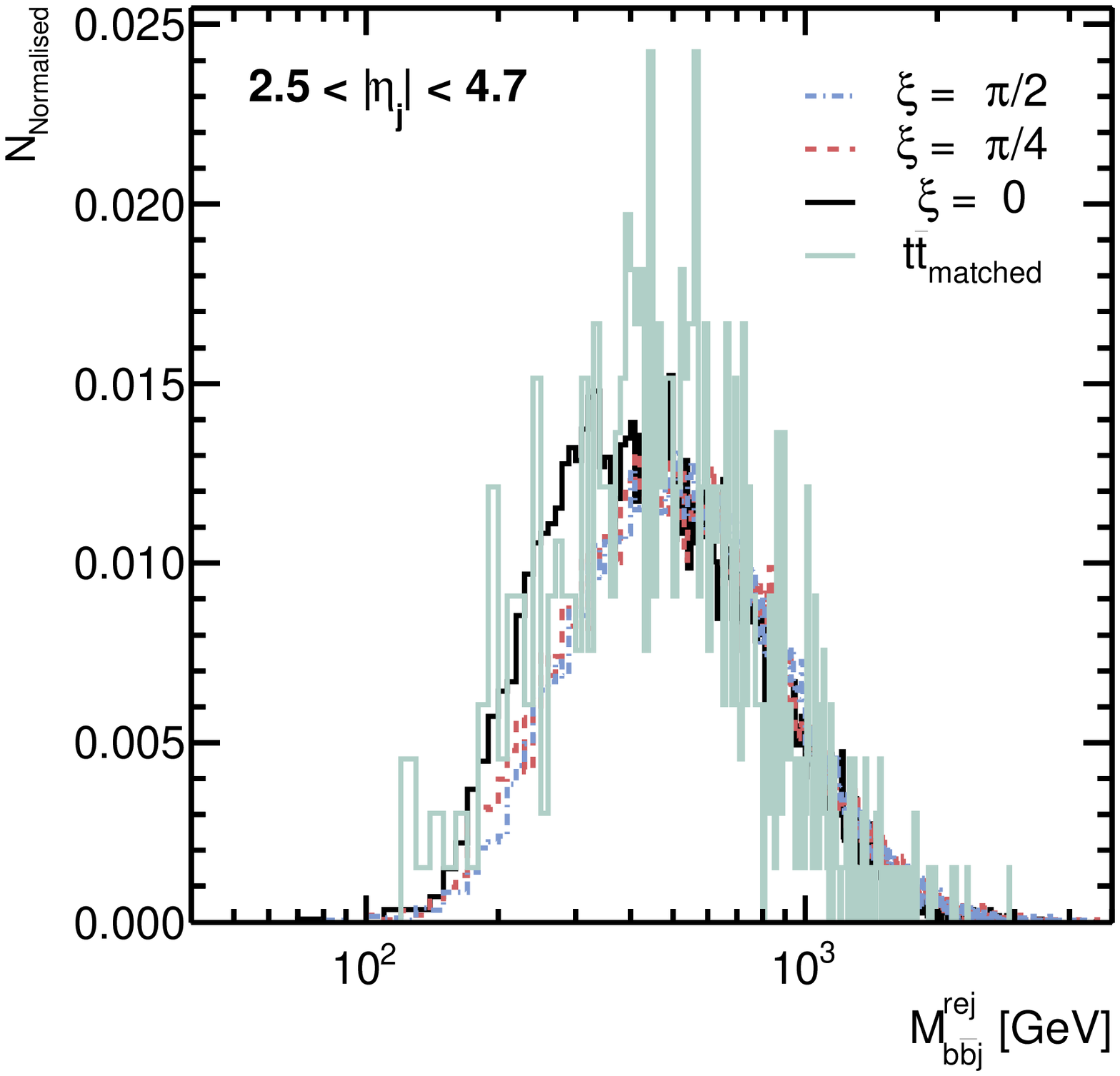}
\includegraphics[width=.45\textwidth,clip=true,trim=6mm 2mm 10mm 10mm]{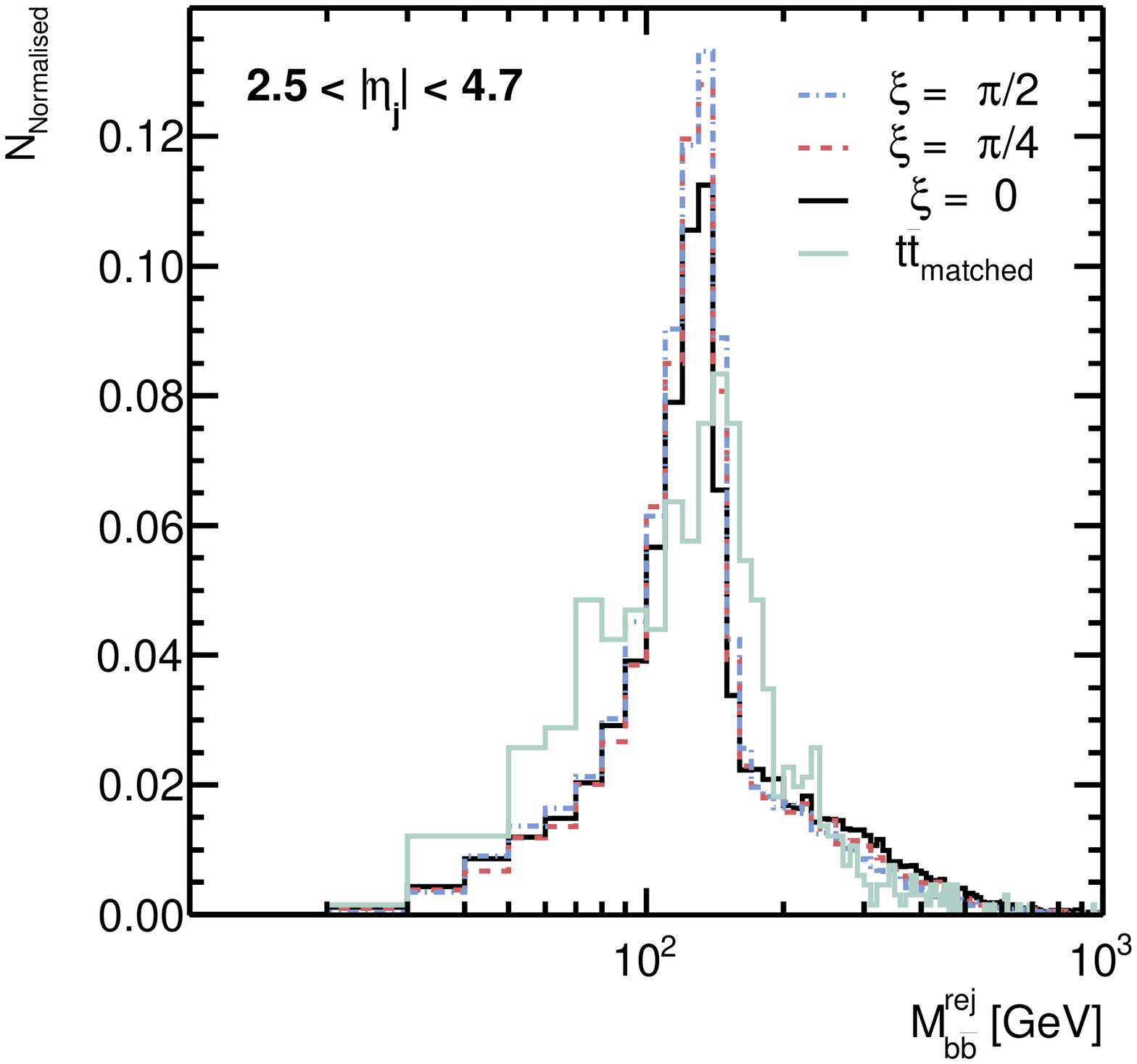}\vspace{-0.5cm}
\caption{Same mass distributions as Fig.~\ref{before}, but with an addition forward jet selection $2.5<|\eta_{j_1}|<4.7$.}
\label{after}
\end{figure}

Similar to Fig.~\ref{before}, we display $M^{rej}_{b\overline{b}j}$ and $M^{rej}_{b\overline{b}}$ distributions with a forward jet cut $2.5<|\eta_{j_1}|<4.7$ in Fig.~\ref{after}. Comparing with Fig.~\ref{before}, we can find that the peak of $M^{rej}_{b\overline{b}j}$ moves towards the high invariant mass region. This indicates that the selected forward jets in $t\overline{t}$  events are not from the top quark hadronic decay. We claim that the $M^{rej}_{b\overline{b}j}$ cut will not be very effective in improving the significance of the signal after the forward jet cut. The $M^{rej}_{b\overline{b}}$ distribution of the backgrounds becomes slightly more flat than the one in Fig.~\ref{before}.

In Tab.~\ref{cutflow}, we summarise the cut-flow cross sections of the signal and backgrounds for 14 TeV LHC. Considering that  the 95\% C.L. from the current LHC data still allows the $\mathcal{CP}$-violating phase $\xi$ to vary within $|\xi| \lesssim0.6 \pi$ (c.f. Tab.~\ref{cpv}), we take three benchmark points $\xi = 0$, $\pi/4$ and $\pi/2$ and assume $y_t =y^{SM}_t$ to demonstrate the observability of our signals. From Tab.~\ref{cutflow}, we can find that: (1) Since there is always a top quark in the signal and backgrounds, the identification procedure of $b$-jet from top quark using $M^{min}_{b\ell}$ reduces both signal and background slightly;
(2) The  {$t\overline{t}$}  events can be further reduced by about one order with the forward jet selection, whilst the signal only loses about $1/3$ events;
(3) The Higgs mass window cut can further remove about $3/4$  of the background  events but keep almost 1/3 of events of the signal.

\begin{table}[ht!]
\fontsize{12pt}{8pt}\selectfont
\begin{center}
\newcolumntype{C}[1]{>{\centering\let\newline\\\arraybackslash\hspace{0pt}}m{#1}}
{\renewcommand{\arraystretch}{1.5}
\begin{tabular}{ C{1.5cm} C{2.75cm} C{2.75cm} |C{1.5cm}C{1.5cm}C{1.5cm} | C{3cm}   }
\cline{1-6}
\hline
&\multicolumn{2}{c|}{\multirow{3}{*}{Cuts}}&\multicolumn{4}{c}{$\sigma$ [fb] } \\\cline{4-7}
&&&\multicolumn{3}{c|}{$thj$} &\multirow{2}{*}{{$t\bar{t}_{\rm matched}$} } \\\cline{4-6}
&&&{$\xi=0$} & {$\xi=\pi/4$} &{$\xi=\pi/2$}      \\\cline{1-7}
\multirow{4}{*}{(C1)} &$ \Delta R_{ij} > 0.4$,  & $i,j = b,j \ \text{or}\  \ell$ & \multirow{4}{*}{0.3169}&\multirow{4}{*}{0.6700}
& \multirow{4}{*}{2.1860}&\multirow{4}{*}{712.4}\\
&   $p_{T}^b > 25 \ \text{GeV}, $     &$|\eta_b|<2.5$&&&&\\
&     $p_{T}^\ell > 25 \ \text{GeV}$,    &$|\eta_\ell|<2.5$&&&&\\
&    $ p_{T}^j > 25 \ \text{GeV}$,    &$|\eta_j|<4.7$&&&&\\    \hline
(C2)&      \multicolumn{2}{ c| }{ $M_{b\ell}<200$ GeV} &0.3152&0.6582&2.1446&708.7\\ \hline
(C3)&      \multicolumn{2}{ c| }{ $ |\eta_j|>2.5$} &0.1492&0.3314&1.1002&80.33\\ \hline
(C4)& \multicolumn{2}{c|}{$ |M_{b_1\overline{b}_2}-m_h|< 15 \ \text{GeV}$}&0.0443&0.1102&0.3762&15.82\\\hline
&  \multicolumn{2}{ c| }{$S/\sqrt{B}$ with 3000 fb$^{-1}$ }&0.610&1.517&5.180\\\hline
&  \multicolumn{2}{ c| }{$S/ B$}&0.28\% & 0.70\% & 2.38\%\\
\hline
\end{tabular}}
\caption{Cutflow of the cross sections (fb) for the signals ($\xi=0,\pi/4$ and $\pi/2$) and the backgrounds at 14 TeV LHC. The conjugate process $pp\rightarrow \overline{t}hj$ has been included. \label{cutflow}}
\end{center}
\end{table}

For each signal point, we calculate the statistical significance $S/\sqrt{B}$ and systematic significance $S/B$ for the luminosity $L = 3000~{\rm fb}^{-1}$. 
From Tab.~\ref{cutflow}, we can see the observability of the pure pseudoscalar interaction ($\xi=\pi/2$)  at 14 TeV HL-LHC is the most promising, with a $\sim 5\sigma$ level statistical significance but a low systematic significance $\sim 2.4\%$. Furthermore, it should be noted that when $\xi=\pi/2$, the values of $y_t/y^{SM}_{t}$ are required to be within the range $0.4 -0.6$ to remain consistent with the current Higgs data. This will  reduce the $thj$  cross section of by a factor $\sim 2/3$,  making the observation more challenging at the LHC.

Since the $\mathcal{CP}$-violating interaction described by (\ref{cpv}) can affect the chirality of $tbW$ coupling through the interference between different Feynman diagrams, we will investigate the top quark polarisation asymmetry in the process $pp \to t(\to \ell^+\nu_\ell b)h(\to b\overline{b})j$. The angular distribution of the lepton from a polarised top quark is given by:
\begin{equation}
   \frac{1}{\Gamma}\frac{d\Gamma}{d\cos\theta_\ell}=\frac{1}{2}(1+\mathcal{P}_t\kappa_\ell\cos\theta_\ell).
\end{equation}
where  the lepton spin analysing power $\kappa_\ell$ is one at tree level in the SM, $\theta_\ell$ is the angle between the direction of the top quark and the lepton momenta in the rest frame of the top quark and $\mathcal{P}_t$ is the spin asymmetry. We reconstruct the top quark rest frame by minimising $\chi^2$, which is defined as:
\begin{equation}
   \chi^2= \bigg( \frac{m_t-m_{\ell\nu_\ell b} }{\Delta {m_t}}\bigg)^2,
\end{equation}
where $\Delta{m_t}$ is taken as the SM top quark decay width \cite{pdg}. With the on-shell condition of the $W$ boson and the top quark, the longitudinal momentum of the neutrino $p_{\nu L}$ can be determined as:
\begin{equation}
  p_{\nu L} = \frac{1}{2p_{\ell T}^2}\bigg( A_{W} p_{ \ell L} \pm E_\ell\sqrt{A_{W}^2 - 4 p_{\ell T}^2  {\not\!\!{E}}^2_T}\bigg),\label{neutrino}
\end{equation}
where $A_W= m_W^2 + 2 \mathbf{p}_{\ell T}\cdot { \not\!\! \mathbf{E}_T}$. The ambiguity of the sign in (\ref{neutrino}) can be removed by the minimal
$\chi^2$ requirement \cite{qinghong}.

\begin{figure}[h!]
\centering
\includegraphics[width=.45\textwidth,clip=true,trim=8mm 6mm 10mm 10mm]{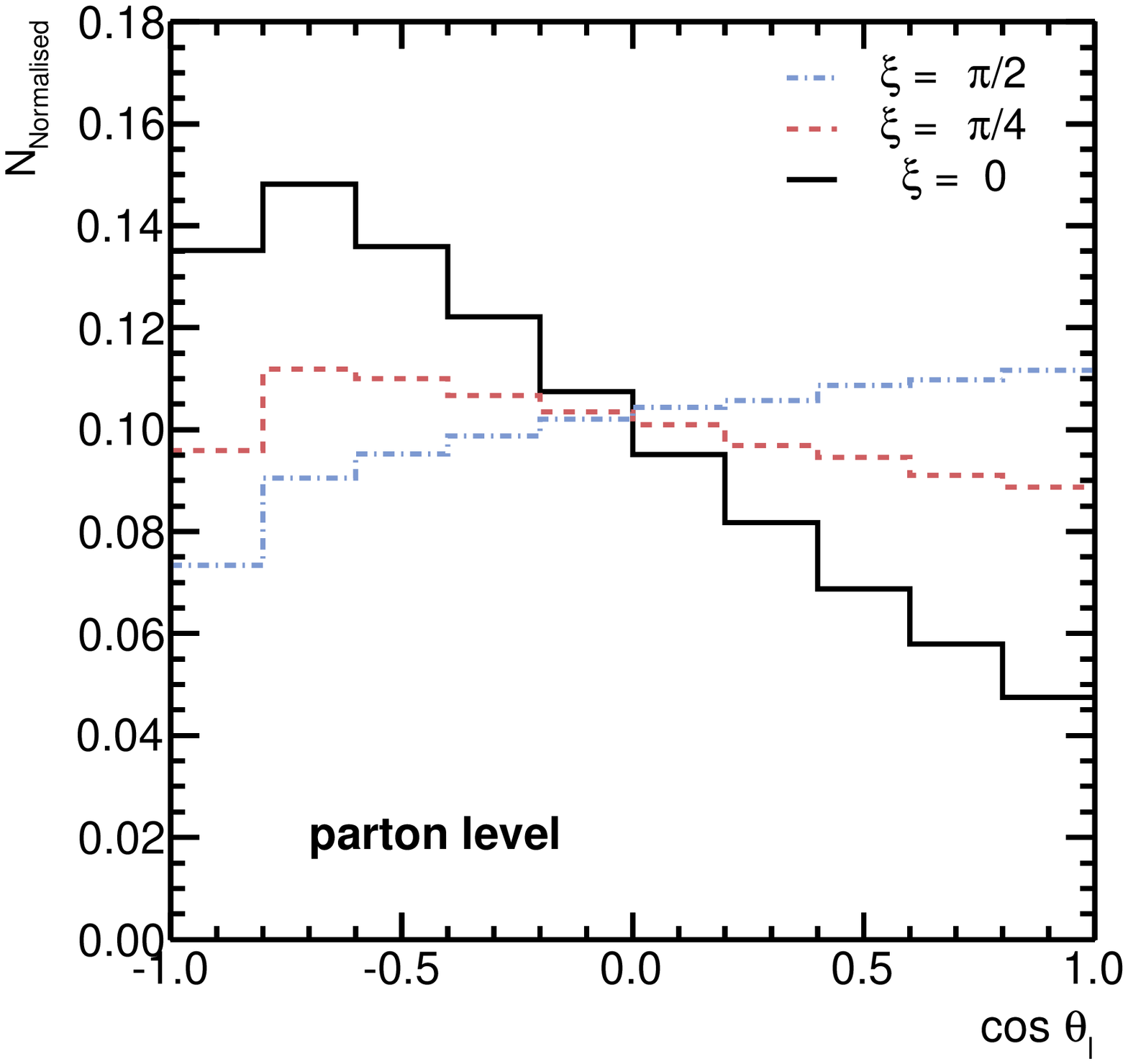}
\includegraphics[width=.45\textwidth,clip=true,trim=8mm 6mm 10mm 10mm]{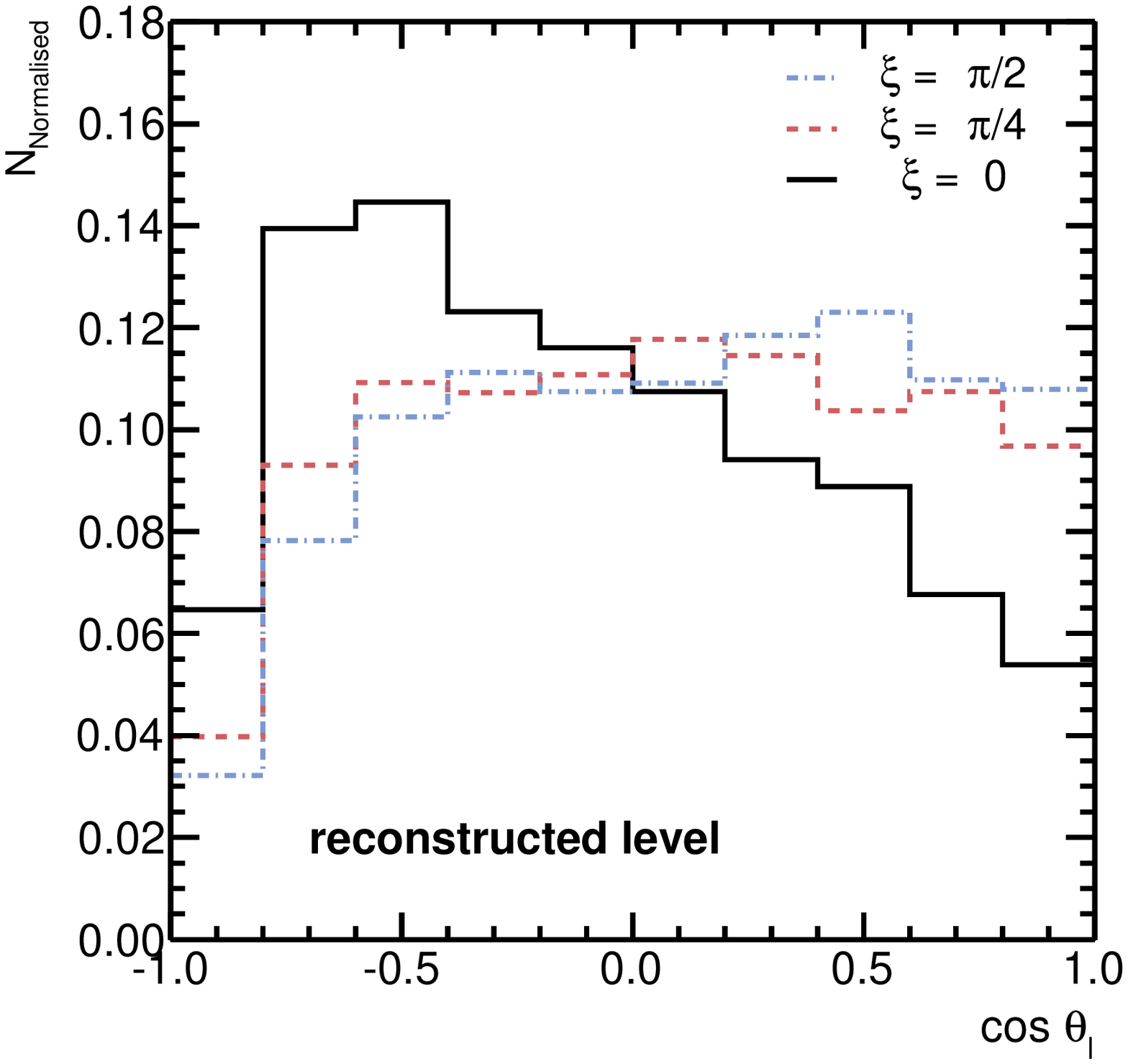}
\caption{The angular distributions of the lepton from the top quark decay in the signal $pp \to t(\to \ell^+\nu_\ell b)h(\to b\overline{b})j$ at parton level (left panel) and reconstruction level (right panel) after the cuts.}
\label{pol}
\end{figure}

In Fig.~\ref{pol}, we plot the lepton angular distributions $\cos\theta_{\ell}$ for the $\mathcal{CP}$ phase $\xi=0, \ \pi/4$ and $\pi/2$ in the signal $pp \to t(\to \ell^+\nu_\ell b)h(\to b\overline{b})j$  at parton level  and reconstruction level respectively after the cuts (C3) and (C4) (c.f. Tab.~\ref{cutflow}). From Fig.~\ref{pol}, we can see that the direction of the lepton is inclined to be opposite to its parent top quark for $\xi=0$, which corresponds to the SM top-Higgs interaction. However, the mixed and pseudoscalar interactions will affect the polarisation state of the top quark and change the slopes such that the number of events with $\cos\theta_\ell > 0$ is larger than those with $\cos\theta_\ell < 0$. The differences between the slopes  are diluted from the parton level to the reconstruction level.

Based on the angular distributions, we can further define the lepton forward-backward (FB) asymmetry:
\begin{equation}
\mathcal{A}^{\ell}_{FB}=\frac{\sigma(\cos\theta_{\ell}>0)-\sigma(\cos\theta_{\ell}<0)}{\sigma(\cos\theta_{\ell}>0)+\sigma(\cos\theta_{\ell}<0)}
\end{equation}
In Tab.~\ref{afb}, we present the values of $A^{\ell}_{FB}$ for $\xi=0,\pi/4$ and $\pi/2$ in the signal {$pp \to t(\to \ell^+\nu_\ell b)h(\to b\overline{b})j$}  at 14 TeV LHC. From Tab.~\ref{afb}, we can see that both of the scalar and pseudoscalar top-Higgs interaction can induce large forward-backward asymmetries but with different signs, reaching -17.6\% and 13.7\% respectively. The measurement of the lepton forward-backward asymmetry can distinguish the scalar and pseudoscalar top-Higgs interactions at the LHC.

\begin{table}[c h!]
\begin{center}
{\renewcommand{\arraystretch}{1.25}
\newcolumntype{C}[1]{>{\centering\let\newline\\\arraybackslash\hspace{0pt}}m{#1}}

\begin{tabular}{  C{1cm} C{3cm}  C{3cm}  C{2cm}}
\hline
 $\xi$& $\sigma(\cos\theta>0)$ [fb] & $\sigma(\cos\theta<0)$ [fb]  & $\mathcal{A}^{\ell}_{FB}$ (\%) \\
\hline
{$0$} &0.01458 &0.0208 &-17.6\\\hline
{$\pi/4$} &0.04687 &0.03991 &8.0\\\hline
{$\pi/2$} &0.1681 &0.1276 &13.7\\\hline
\hline
\end{tabular}}
\caption{The reconstructed-level forward-backward asymmetry $\mathcal{A}^{\ell}_{FB}$ at 14 TeV LHC.}\label{afb}
\end{center}
\end{table}

\section{\label{sec:level1}conclusion}

In this paper, we have obtained constraints on the $\mathcal{CP}$-violating top-Higgs couplings using the current Higgs data and found that values of $\mathcal{CP}$-violating phase $|\xi| > 0.6\pi$ are already excluded at  95\% C.L.. We  expected TLEP to improve this exclusion region to $|\xi |>0.07\pi$. With current constraints on $\xi$, we further investigate the observability of the scalar, pseudoscalar and mixed top-Higgs interactions through the channel $pp \to t(\to \ell^+\nu_\ell b)h(\to b\overline{b})j$. We found that it is most promising to observe pure pseudoscalar interactions at the HL-LHC but it is still challenging due to a low $S/B$ ratio. However, the anomalous top-Higgs couplings can lead to sizeable differences in forward-backward asymmetries and can be distinguished by measuring the lepton angular distributions from polarised top quarks at 14 TeV LHC.

\section*{Acknowledgement}
This work was supported by the Australian Research Council. Lei Wu is also supported by the National Natural Science Foundation of China (NNSFC) under grants Nos. 11222548, 11275057 and 11305049, by Specialised Research Fund for the Doctoral Program of Higher Education under Grant No. 20134104120002.

\end{document}